\def\year{2022}\relax
\documentclass[letterpaper]{article} 
\usepackage{aaai22}  
\usepackage{times}  
\usepackage{helvet}  
\usepackage{courier}  
\usepackage[hyphens]{url}  
\usepackage{graphicx} 
\usepackage[square, numbers]{natbib}  
\usepackage{caption} 
\DeclareCaptionStyle{ruled}{labelfont=normalfont,labelsep=colon,strut=off} 
\usepackage{algorithm}
\usepackage{algorithmic}

\usepackage{newfloat}
\usepackage{listings}
\floatstyle{ruled}
\newfloat{listing}{tb}{lst}{}
\floatname{listing}{Listing}
\title{Benchmarking Differentially Private Synthetic Data Generation Algorithms\thanks{Work done while at Tumult Labs.}}
\author{
    Yuchao Tao, \textsuperscript{\rm 1,3}
    Ryan McKenna, \textsuperscript{\rm 1,2}\\
    Michael Hay, \textsuperscript{\rm 1,4}
    Ashwin Machanavajjhala, \textsuperscript{\rm 1,3}
    Gerome Miklau \textsuperscript{\rm 1,2}
}
\affiliations{
    \textsuperscript{\rm 1} Tumult Labs, USA, \ \ 
    \textsuperscript{\rm 2} University of Massachusetts Amherst, USA\\
    \textsuperscript{\rm 3} Duke University, USA, \ \ 
    \textsuperscript{\rm 4} Colgate University, USA\\
    science@tmlt.io
}

\usepackage{amsmath}
\usepackage{graphicx}
\usepackage[colorlinks=true, allcolors=blue]{hyperref}
\usepackage{float}
\usepackage{makecell}
\usepackage{tablefootnote}
\usepackage{caption}
\usepackage{subcaption}
\usepackage{bm}
\usepackage[capitalize]{cleveref}
\usepackage{adjustbox}

\usepackage[colorinlistoftodos]{todonotes}

\newcommand{\mh}[2][]{}
\newcommand{\am}[2][]{}
\newcommand{\yt}[2][]{}

\newcommand{\dataset}[1]{\textit{#1}}
\newcommand{\metric}[1]{\textit{#1}}
\newcommand{\algo}[1]{\texttt{#1}}
\newcommand{\finding}[2]{\textit{F#1: #2}}

\newcommand{\eat}[1]{}

\begin{document}

\maketitle

\begin{abstract}
This work presents a systematic benchmark of differentially private synthetic data generation algorithms that can generate tabular data. 
Utility of the synthetic data is evaluated by measuring whether the synthetic data preserve the distribution of individual and pairs of attributes, pairwise correlation as well as on the accuracy of an ML classification model. In a comprehensive empirical evaluation we identify the top performing algorithms and those that consistently fail to beat baseline approaches. 
\end{abstract}

\section{Introduction}

While there are many compelling reasons to share data about individuals, such sharing is often prevented due to privacy concerns.
Differentially private synthetic data generation stands out as an appealing solution to this problem: it provides strong formal privacy guarantees, while producing a synthetic data set that ``looks like'' the real data from the perspective of an analyst.  This problem has received considerable attention from the research community, with a wide variety of approaches available in the literature.  \cite{DBLP:journals/corr/abs-2108-04978, 
mckenna2019graphical, 
DBLP:conf/sigmod/ZhangCPSX14, zhang2017privbayes, 
DBLP:journals/corr/abs-1802-06739, 
rosenblatt2020differentially, 
DBLP:conf/iclr/JordonYS19a, 
rosenblatt2020differentially, 
DBLP:journals/pvldb/GeMHI21, 
DBLP:conf/icml/VietriTBSW20, 
DBLP:conf/icml/AydoreBKKM0S21, 
DBLP:journals/popets/ChanyaswadLM19, 
gretel,
DBLP:journals/corr/abs-1809-09087,
DBLP:journals/pvldb/LiXZJ14,
DBLP:journals/corr/abs-2106-12949,
DBLP:journals/corr/abs-2012-15128,
DBLP:conf/icml/LiuV0UW21,
DBLP:journals/corr/abs-2106-07153,
DBLP:conf/psd/SnokeS18,
DBLP:conf/aistats/HarderAP21,
DBLP:conf/cvpr/Torkzadehmahani19,
neurips20chen}.   

Despite the variety of mechanisms available for this task, the community is lacking a systematic empirical study that compares a variety of mechanisms on different datasets, tasks, and privacy levels.  Prior work in this space includes \cite{rosenblatt2020differentially}, which focuses on GAN-based algorithms in terms of the machine learning classification accuracy;  \cite{DBLP:journals/pvldb/FanLLCSD20}, which focuses on GAN-based algorithms in terms of the utility of classification, clustering, aggregation queries and privacy protection; \cite{bowen2019comparative}, which focuses on algorithms from the NIST 19 synthetic data challenge in terms of the marginal distribution, joint distribution and correlation;  \cite{bowen2020comparative}, which focuses on algorithms before 2016 in terms of the statistical utility; \cite{DBLP:journals/corr/abs-2004-07740}, which proposes a general framework for evaluating the quality of private synthetic data; and \cite{xu2019modeling}, which proposes a framework SDGym to benchmark the performance of synthetic data. However, none of these works both include a representative set of state-of-the-art algorithms and cover a representative set of metrics.

Inspired by DPBench \cite{DBLP:conf/sigmod/HayMMCZ16}, we focus on benchmarking differentially private synthetic data generation algorithms selected from a specific set of inclusion criteria.  
Most DP synthetic data generation algorithms learn a model over the data from which synthetic data records are sampled. We categorize the algorithms included in our study into three broad classes: \textbf{GAN-based} methods learn a generative adversarial network (GAN) privately, mainly by adding noise to the gradient calculation; \textbf{Marginal-based} methods measure a subset of the low-order marginals and use them to fit a graphical model; and \textbf{Workload-based} algorithms iteratively improve their model to reduce approximation error on workload queries.

We evaluate these mechanisms across different datasets and privacy budgets on whether the synthetic data can preserve the distribution of individual and pairs of attributes, pairwise correlation and on the accuracy of an ML classification model. Our experiments reveal a number of findings:
\begin{enumerate}
    \itemsep0em 
    \item Many mechanisms, especially GAN-based mechanisms, often fail to preserve the most basic statistics of the data distribution --- their one way marginals.  Moreover, these mechanisms fail to beat simple baseline mechanisms on other more interesting metrics. 
    \item No single mechanism is best on every dataset and task, and privacy budget considered.  However, marginal-based mechanisms consistently rank among the best.
    \item Marginal-based methods expect discrete data, and proper discretization is essential to get good performance on numerical attributes.  We found that using PrivTree \cite{zhang2016privtree} to discretize numerical attributes is far more effective than equal-width discretization. 
\end{enumerate}

\section{Methodology}

In this section, we describe the mechanisms included in this study (and the justification for inclusion), the tasks considered, the datasets evaluated, as well as any modifications necessary to run the mechanisms on our datasets. 

\subsection{Mechanisms}
We consider five inclusion criteria for selecting the mechanisms for this benchmark study, enumerated below: 

\begin{enumerate}
    \item \textbf{End-to-End DP}: It is claimed to be an end-to-end differential private algorithm that takes a tabular dataset as input and generates a synthetic data of the same schema.
    \item \textbf{Tabular Data}: It supports tabular data that could have numerical and/or categorical columns. The associated publication includes experiments on tabular data.
    \item \textbf{Publication Venue}: It is published in a top conference/journal or included in a well known library. For example, we consider academic venues of SIGMOD, VLDB, CCS, NeurIPS, ICML, PETS and JPC and the open-source libraries of SmartNoise and Gretel. Algorithms from other conferences/journals and libraries are left for future work.
    \item \textbf{Publicly Available Source Code}: Its source code is accessible to the public (e.g. either available on GitHub or linked in the paper describing the work).
    \item \textbf{No Public Data}: It requires no public data.
\end{enumerate}

Table \ref{tab:inclusion} lists the mechanisms included in this benchmark, categorizing them by type. 
\mh{clarify that the interface does not let you set the budget a priori}
\mh{explain what empty data set means: i.e., its generator generator invalid records}
Included in this table is \algo{GretelRNN}, which satisfied the inclusion criteria but is not shown in the experimental results because we found that even when the epsilon that it claims is over one million, it generates an empty dataset after one hour of the generation stage due to the high rejection rate of invalid samples.

\begin{table}[h!]
    \centering
    {\small 
    \begin{tabular}{|p{2.8cm}|p{1cm}|p{1.5cm}|}
    \hline
        \textbf{Algorithm} & 
        \textbf{Code} & 
        \textbf{Type} \\ \hline

        \algo{MST} \cite{DBLP:journals/corr/abs-2108-04978} & \cite{code-mst} & Marginal \\ \hline
        \algo{MWEM-PGM} \cite{mckenna2019graphical} & \cite{code-mwempgm} & Marginal \\ \hline
        \algo{PrivBayes} \cite{DBLP:conf/sigmod/ZhangCPSX14, zhang2017privbayes} & \cite{code-privbayes} & Marginal \\ \hline
        \algo{DPGAN} \cite{DBLP:journals/corr/abs-1802-06739} & \cite{code-smartnoise} & GAN \\ \hline
        \algo{DPCTGAN} \cite{rosenblatt2020differentially} & \cite{code-smartnoise} & GAN \\ \hline
        \algo{PATEGAN} \cite{DBLP:conf/iclr/JordonYS19a} & \cite{code-smartnoise} & GAN \\ \hline
        \algo{PATECTGAN} \cite{rosenblatt2020differentially} & \cite{code-smartnoise} & GAN \\ \hline
        \algo{FEM} \cite{DBLP:conf/icml/VietriTBSW20} & \cite{code-fem} & Workload \\ \hline
        \algo{RAP} \cite{DBLP:conf/icml/AydoreBKKM0S21} & \cite{code-rap} & Workload \\ \hline
        \algo{Kamino} \cite{DBLP:journals/pvldb/GeMHI21} & \cite{code-kamino} & Other \\ \hline
        \algo{RON-GAUSS} \cite{DBLP:journals/popets/ChanyaswadLM19} & \cite{code-rongauss} & Other \\ \hline
        \algo{GretelRNN} \cite{gretel} & \cite{gretel} & Other \\ \hline

    \end{tabular}
    \caption{Mechanisms included in our study.}
    \label{tab:inclusion}
    }
\end{table}
While we expect all the mechanisms to be able to take a dataset with mixed-type columns as input, some of them only accept categorical datasets or numerical datasets. For mechanisms that expect numerical data, we one-hot encode all categorical features.  For mechanisms that expect categorical data, we discretize all numerical features.  We considered two approaches for discretization: an equal-width binning strategy, and a strategy based on PrivTree \cite{zhang2016privtree}. We find that PrivTree binning was never  worse than equi-width binning in most of the cases and lead to significant improvements for some metrics.  Details omitted due to space.

Three mechanisms, \algo{MWEM-PGM}, \algo{FEM} and \algo{RAP}, also require a workload as input.  For datasets that include a classification label (see next section), we set the workload to be all 2- and 3-way marginals that include the label as one of the attributes.  For other datasets, we set the workload to be all 2-way marginals. For all algorithms except Kamino, we use default hyper-parameters. For Kamino, the search algorithm (Algorithm 6 from~\cite{DBLP:journals/pvldb/GeMHI21}) was not included in the available implementation, so we implemented a variant of it.



\subsection{Datasets}
We consider seven datasets with different characteristics, numbers of records, and column types. Datasets \dataset{Car} and \dataset{Mushroom} contain only categorical attributes; \dataset{Scooter} contains only numerical attributes; all other datasets contain a mix of attribute types. Most datasets have a classification label.  All datasets are from the UCI machine learning repository \cite{Dua:2019} except \dataset{Scooter} which is from Gretel \cite{scooter}.
\begin{table}[H]
    \centering
    \begin{tabular}{c|p{1cm}p{1cm}p{1cm}p{1cm}}
    Name & Records & Cat. & Numeric & Label \\\hline
    Shopping & 12330 & 9 & 10 & Yes \\
    Adult & 32561 & 9 & 6 & Yes \\
    Bank & 45211 & 13 & 8 & Yes \\
    Census & 299285 & 29 & 12 & Yes \\
    Car & 1728 & 7 & 0 & Yes \\
    Mushroom & 8124 &  23 & 0 & Yes \\
    Scooter & 27715 & 0 & 5 & No \\
    \end{tabular}
    \caption{Summary of datasets.}
    \label{tab:datasets}
\end{table}

\subsection{Metrics}
\label{sec:metrics}
We consider four groups of metrics to measure the goodness-of-fit of the synthetic data generated by each algorithm. Each group might include more than one metric but with similar goals.\footnote{For brevity, we include a single metric per group. In the full version of this paper, multiple metrics per group are considered.} These metrics are inspired by SDGym \cite{sdgym}. 
For the first three metric groups, numerical attributes are discretized into 19 bins of equal-depth (based on the original data). Since the algorithms may generate synthetic data that lies outside of this range, an additional bin is added to each end of the range.
\begin{enumerate}
    \item \textbf{Individual Attribute Distribution Similarity (Ind)} 
    This group of metrics measures the similarity of one-way marginals between the synthetic data and the original data. We use total variation distance (TVD), to measure the distance between two one-dimensional distributions, and use 1-TVD as the score. We report the average score over all one-way marginal as the final score.  
    \item \textbf{Pairwise Attribute Distribution Similarity (Pair)} Similar to the Individual Distribution Similarity, this group of metrics measures the TVD for each two-way marginal and we average across all attribute pairs. 
    \item \textbf{Pairwise Correlation Similarity (Corr)}  We use Cramer's V with bias correction \cite{bergsma2013bias} to measure the correlation between two attributes, and, following convention, discretize it into one of four levels: V in [0, .1) is low, [.1, .3) is weak, [.3, .5) is middle and [.5, 1) is strong. 
    The metric \metric{CorAcc} measures the accuracy of correlation levels, reporting the fraction of pairs where the synthetic and original data assign the same correlation level.
    \item \textbf{Classification Accuracy (F1)} We use the synthetic dataset to train an XGBoost classifier and use it to make predictions on the original data. The score is reported by the f1 score using macro average. This metric category only applies to datasets that have a class label.
\end{enumerate}

\eat{
\subsection{PrivTree Discretization}
For algorithms that need discretization for each single numerical column as an algorithm repair step, we consider two options: equal-width binning with 100 bins and PrivTree binning \cite{zhang2016privtree}. PrivTree binning uses PrivTree to partition the 1-D domain of a single attribute as bins. We use the default parameters from the original paper. When PrivTree binning is used, we allocate a fixed amount privacy budget 0.1 for PrivTree binning, and leave the rest for the private data generation algorithm. If the total privacy budget is below 0.1, we instead use the 20 percent of the privacy budget for PrivTree binning.
}

\section{Findings}



\eat{
In this section, we analyze the scores measured by different metrics for the synthetic datasets generated from each private data generation mechanisms. 
Mechanisms with a plus sign in the name indicating that it uses PrivTree as the discretization strategy.\mh{I think we should consider dropping the + sign.  It seems weird to call out this ``algorithm repair'' and not others.} Our experiments show that using PrivTree is superior than equal-width binning in terms of the scores, so by default we use PrivTree for discretization.\mh{doesn't this belong in algorithm repair?} 
\algo{GretelRNN} is removed from the benchmark analysis because it cannot generate the dataset with high rejection rate for each sample. All mechanisms are repeated 5 times for each combination of 7 datasets and 3 epsilons.
}


We use SDGym \cite{xu2019modeling} as the platform for all the experiments. The privacy parameter $\epsilon$ varies within $\{0.1, 1.0, 10.0 \}$. In this section, we briefly summarize our main findings.  

\begin{figure}[!t]
    \centering
    \begin{subfigure}[t]{0.1535\linewidth}
        \centering
        \includegraphics[width=\textwidth]{"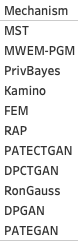"}
    \end{subfigure}
    \begin{subfigure}[t]{0.405\linewidth}
        \centering
        \includegraphics[width=\textwidth]{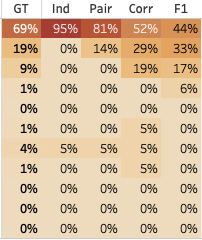}
        \caption{\textbf{Optimal Rate} }
        \label{fig:optimal}
    \end{subfigure}
    \begin{subfigure}[t]{0.405\linewidth}
        \centering
        \includegraphics[width=\textwidth]{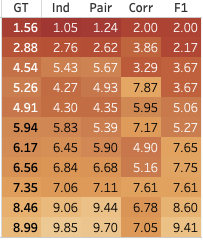}
        \caption{\textbf{Average Ranking} }
        \label{fig:ranking}
    \end{subfigure}
    \caption{Overview of mechanisms in terms of optimal rate and average ranking across datasets, epsilons, and metrics stratified by metric groups. GT means ``Grand Total.''}
    \label{fig:overview}
\end{figure}

\finding{1}{No algorithm dominates}.  We consider a mechanism ``optimal'' for a particular combination of dataset, epsilon, and metric if that mechanism achieves the best performance (averaged over trials) according to the metric. The optimal rate, shown in \cref{fig:optimal}, is the frequency at which a mechanism is optimal for a particular category of metric.  Any algorithm that has a non-zero optimal rate means that the algorithm performs the best on at least one combination of dataset, epsilon, and metric.  Over half of the algorithms have a non-zero optimal rate.

\finding{2}{While no algorithm dominates, marginal-based approaches are highly ranked and \algo{MST}, in particular, is the top-ranked algorithm across all metrics}. 
To get a sense of the overall best performing algorithm, we rank the algorithms according to each metric and then average the rankings.  In~\cref{fig:ranking}, we report the average ranking, stratified by category of metric; we also report the average ranking across all metrics (``GT'').  The overall average rank of \algo{MST} is 1.56 indicating that it is frequently the best algorithm, which is also consistent with the results of~\cref{fig:optimal}.

\begin{figure}
    \centering
    \begin{subfigure}[t]{\linewidth}
        \centering
        \includegraphics[width=\linewidth]{"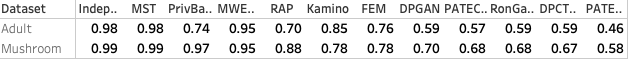"}
        \caption{1-TVD for individual attribute distributions}
        \label{fig:ind_l1}
    \end{subfigure}
    \begin{subfigure}[t]{\linewidth}
        \centering
        \includegraphics[width=\linewidth]{"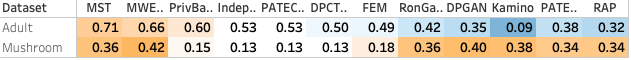"}
        \caption{Correlation accuracy (\metric{CorAcc})}
        \label{fig:CorAcc}
    \end{subfigure}
    \begin{subfigure}[t]{\linewidth}
        \centering
        \includegraphics[width=\linewidth]{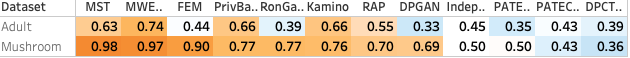}
        \caption{Classification accuracy, measured by f1 score, of an XGBoost classifier. The F1 score by training on the original data is 0.86 for Adult and 1.0 for Mushroom. }
        \label{fig:XGBoost}
    \end{subfigure}
    \caption{Performance metrics for synthetic data algorithms at $\epsilon=1.0$}
\end{figure}

\finding{3}{Many mechanisms fail to accurately preserve the distributions of individual attributes (1-way marginals).} \cref{fig:ind_l1} reports the average similarity (1-TVD) of individual attribute distributions for two of the datasets in our benchmark, \dataset{Adult} and \dataset{Mushroom}.  Several algorithms have an average similarity of less than 0.75.  \algo{PrivBayes} has uneven performance, doing well on \dataset{Mushroom} and worse on \dataset{Adult}; we hypothesize this is due to how PrivBayes discretizes numerical attribues (\dataset{Mushroom} has no numerical attributes).

\begin{figure}
    \centering
    \begin{subfigure}[t]{0.95\linewidth}
        \centering
        \includegraphics[width=\textwidth, trim={10 30 10 20},clip]{"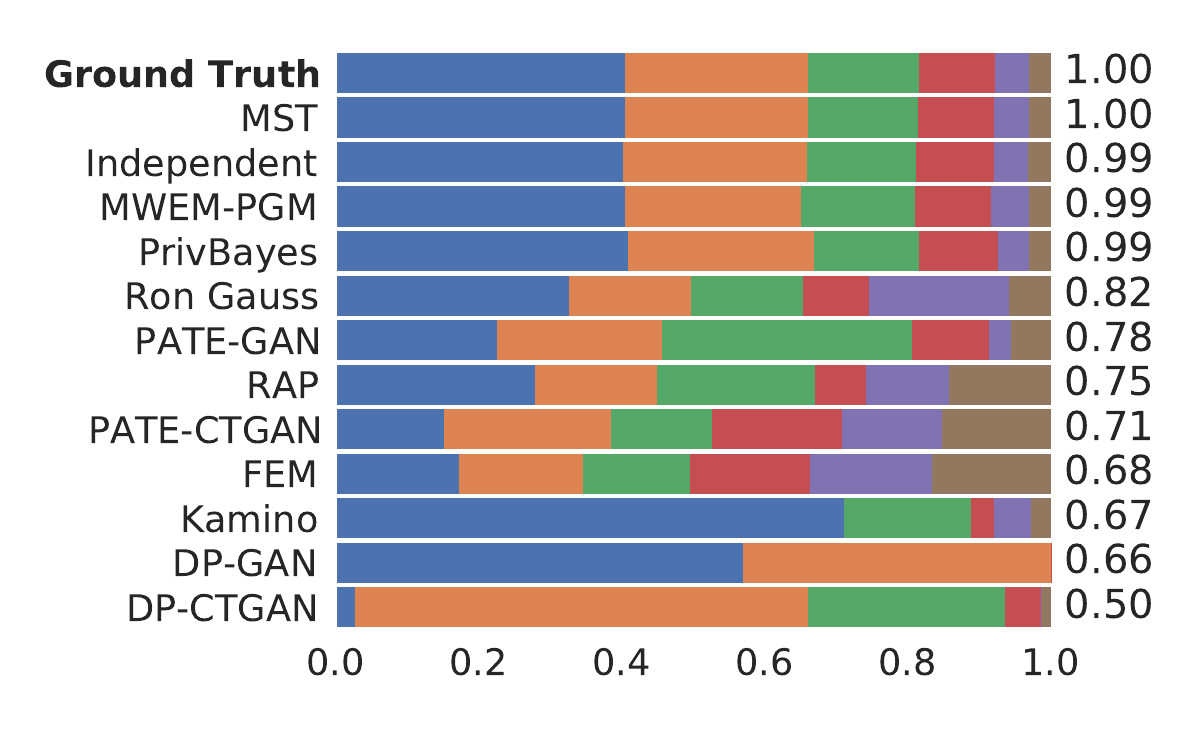"}
        \caption{Distributions of the \emph{relationship} attribute; a categorical attribute with six possible values (shown as different colors). \label{fig:relationship_1way}}
    \end{subfigure}
    \begin{subfigure}[t]{0.95\linewidth}
        \centering
        \includegraphics[width=\textwidth, trim={10 30 10 20},clip]{"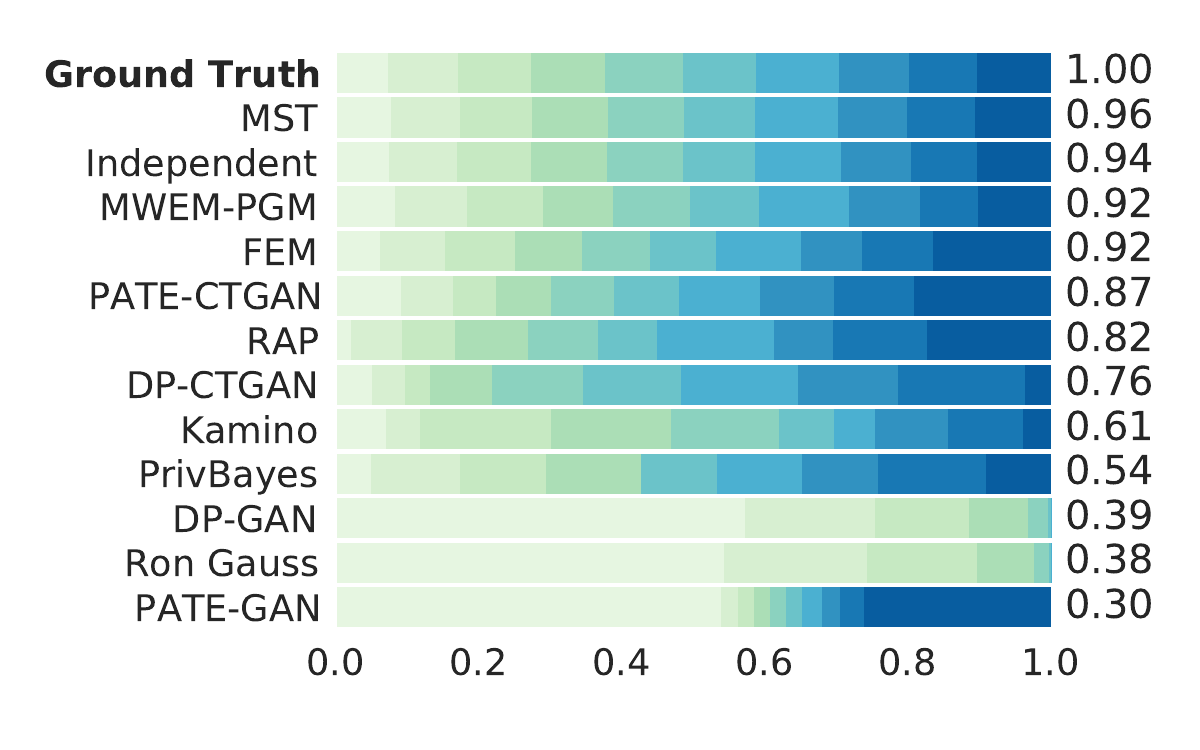"}
        \caption{Distributions of the \emph{age} attribute; a numerical attribute discretized by quantiles (which is why the \emph{Ground Truth} appears uniform).
        \label{fig:age_1way}}
    \end{subfigure}
    \caption{One-way marginal distributions for the original \dataset{Adult} dataset and
    for a sample synthetic dataset generated by each algorithm ($\epsilon=1.0$) in descending order by similarity (1-TVD, shown to the right of each row).
     \label{fig:attribute_distributions}}
\end{figure}

To gain some intuition for how well algorithms are preserving attribute distributions, we display some representative examples in~\cref{fig:attribute_distributions} from the \dataset{Adult} dataset.  \cref{fig:relationship_1way} uses a stacked bar chart to compactly display the distribution of the \emph{relationship} attribute.  The first row is the distribution in the original data (ground truth) and the remaining rows are the distributions in the synthetic datasets produced by the algorithms, ordered by their similarity to the ground truth (1-TVD is reported to the right of each row).  When 1-TVD is below 0.75, the distortion is visually apparent.  Some algorithms (\algo{DP-GAN}, \algo{DP-CTGAN}) have highly skewed distributions; others appear uniform (\algo{FEM}) even though the original data is non-uniform.  \cref{fig:age_1way} shows the distributions of the numerical attribute \emph{age} (after it was discretized).\mh{say something about results on age.} \yt{PrivBayes actually has prob 0 for one bin, which might explains why it has a low score. }

In addition to looking at individual attribution distributions, we also evaluate pairwise attribute correlations.  \cref{fig:CorAcc} reveals the extent to which correlations are accurately preserved.  It reports the \metric{CoreAcc} metric for datasets~\dataset{Adult} and \dataset{Mushroom}.  As a baseline for comparison, we include \algo{Independent}, an algorithm that assumes all columns are statistically independent (uncorrelated) and generates synthetic data by sampling attribute values from distributions estimated from 1-way marginals perturbed with Laplace noise. We use a divergent color scheme to indicate whether it is above (orange) or below (blue) the baseline.

The results in \cref{fig:CorAcc} give us two main findings.
\finding{4}{In terms of preserving attribute correlations, Marginal-based algorithms consistently obtain the highest correlation accuracy.}\mh{An exception is PrivBayes on Mushroom, which I cannot explain.}\yt{Interesting. We need to look at after today.} And, \finding{5}{Many algorithms fail to preserve correlations more accurately than independent, a simple baseline that generates uncorrelated data}.

\begin{figure}[t]
    \centering
    \includegraphics[width=0.9\linewidth, trim={80 200 80 170}, clip]{"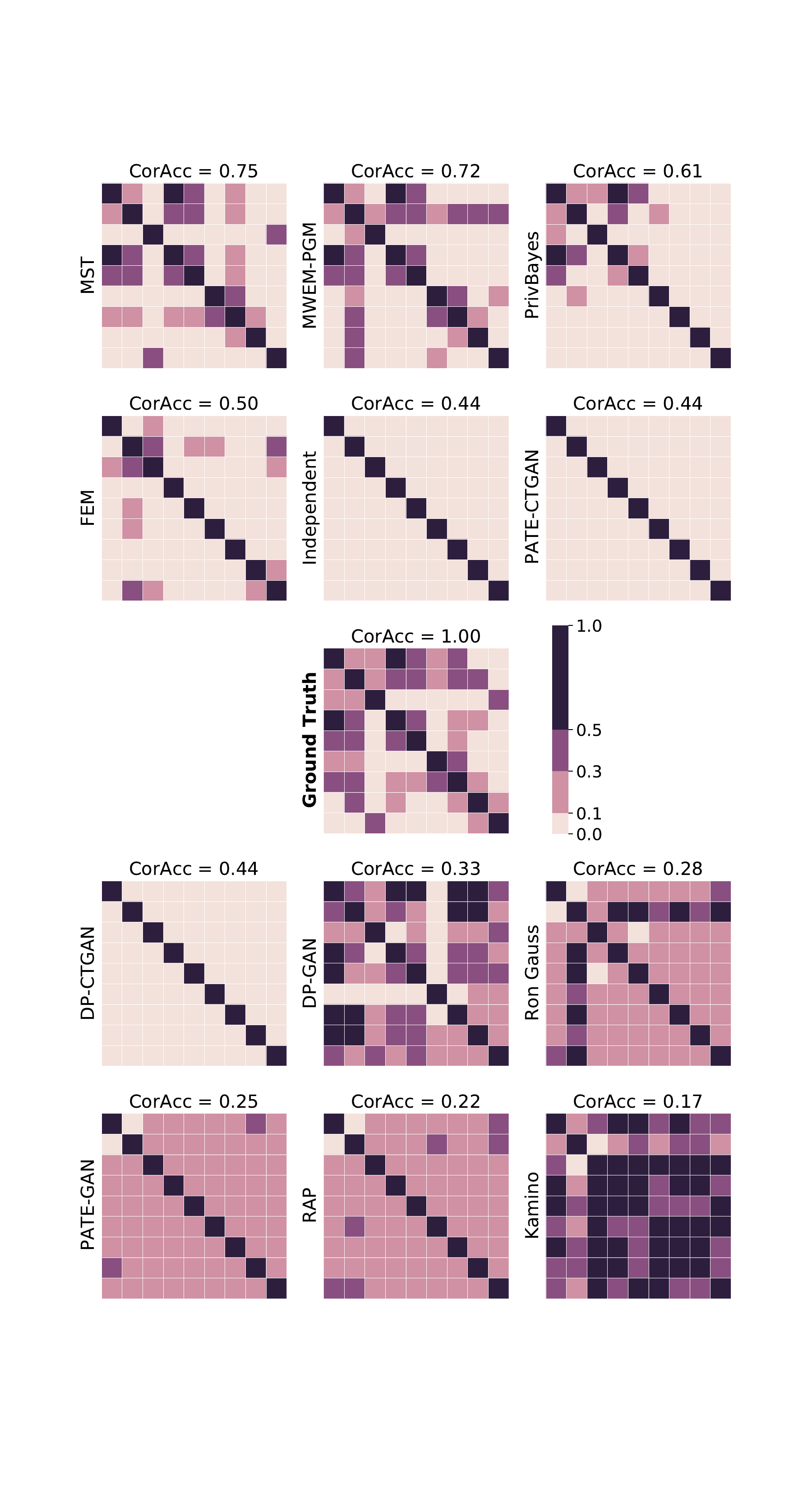"}
    \caption{Correlation heatmaps for all pairwise categorical attributes from the \dataset{Adult} dataset.  A heatmap is shown for the original data, \emph{Ground Truth} (center), and for a sample synthetic dataset generated by each algorithm at $\epsilon=1.0$.
    Attributes are sorted by domain sizes.
    }
    \label{fig:correlation}
\end{figure}

To gain some intuition about correlations, we look more closely at the correlation accuracy on the \dataset{Adult} dataset. \cref{fig:correlation} shows correlation heatmaps for the original data (\dataset{Ground Truth}, center plot) and for synthetic datasets generated by the algorithms.  In each heatmap, a cell corresponds to an attribute pair and darker cells indicate stronger correlation (the colors are discretized to the four correlation levels described earlier).   The figure shows that marginal-based algorithms (top row) do fairly well (\metric{CorAcc}=0.75 means 75\% of the colored cells match the ground truth figure) though some correlations are not captured.  Several algorithms (\algo{FEM}, \algo{PATE-CTGAN}, \algo{DP-CTGAN}) have accuracy matching the baseline \algo{Independent}.  The correlation plots show why: the synthetic data generated by these algorithms has attributes that appear to be statistically independent (uncorrelated), matching the independent baseline.  The remaining algorithms have accuracy that is \emph{lower} than the baseline and it appears that this is due to those techniques introducing \emph{spurious} correlation.

\finding{6}{The synthetic data produced by marginal-based approaches \algo{MST} and \algo{MWEM-PGM} is of sufficient quality that it can be used to train an accurate classifier, nearly matching the performance of a classifier trained on the original data.}
In \cref{fig:XGBoost}, we report how well the synthetic data preserves the ability to train a classifier.  It shows the f1 score of an XGBoost classifier trained on the synthetic data.  On \dataset{Mushroom}, the classifiers trained on the synthetic data from \algo{MST} and \algo{MWEM-PGM} achieve nearly perfect accuracy; on \dataset{Adult}, \algo{MWEM-PGM} achieves the highest f1 score of 0.74, which approaches the f1 on the original data of 0.86.  The relatively strong performance of \algo{MWEM-PGM} may reflect the fact that its strategy is tuned to support classifier learning by favoring marginals that include the class label.

\finding{7}{The synthetic data produced by GAN-based approaches yields classifers that are generally no more accurate than a simple majority classifier.}  In~\cref{fig:XGBoost}, we include baseline algorithm \algo{Independent}.  Since this algorithm models each attribute independently, a classifier trained on its synthetic data can be no more accurate than a classifier that always predicts the majority label.  In this figure, we again use a divergent color scheme to compare performance to this basline and we see the GAN-based approaches often have an f1 score below the baseline.  

\section{Conclusion}
We presented a systematic benchmark study of differentially private synthetic data generation algorithms that can generate tabular data. We considered a variety of algorithms including GAN-based, Marginal-based and Workload-based methods and evaluated their utility in terms of how well they preserve low dimensional statistics, pairwise correlations and ML classification accuracy. We found that Marginal-based methods consistently outperformed other methods, and GAN-based methods were unable to preserve the 1-dimensional statistics of tabular data. Our research motivates future research directions that include developing better GAN methods for tabular data, methods for pre-processing categorical/numeric data types, and identifying methods to choose the best synthetic data algorithms given a dataset.  

\clearpage
\bibliographystyle{abbrvnat}
\bibliography{refs}

\end{document}